%% file: main.tex
\documentclass[sigconf]{acmart}
\usepackage{algorithm}
\usepackage{algorithmic}
\usepackage{xspace}
\usepackage{booktabs}
\usepackage{tabularx} 
\usepackage{multirow}

\newcommand{\A}{$\mathbf{A}$\xspace}
\newcommand{\B}{$\mathbf{B}$\xspace}
\newcommand{\C}{$\mathbf{C}$\xspace}
\AtBeginDocument{%
  }

\copyrightyear{2025}
\acmYear{2025}
\setcopyright{acmlicensed}\acmConference[ICPE '25]{Proceedings of the 16th ACM/SPEC International Conference on Performance Engineering}{May 5--9, 2025}{Toronto, ON, Canada}
\acmBooktitle{Proceedings of the 16th ACM/SPEC International Conference on Performance Engineering (ICPE '25), May 5--9, 2025, Toronto, ON, Canada}
\acmDOI{10.1145/3676151.3719365}
\acmISBN{979-8-4007-1073-5/2025/05}

\usepackage{graphicx}
\usepackage{adjustbox}
\usepackage{caption}
\begin{document}

\title{Parallel GPU-Enabled Algorithms for SpGEMM on Arbitrary Semirings with Hybrid Communication}

\author{Thomas McFarland}
\email{tfm62@cornell.edu}
\orcid{0009-0007-3349-3066}
\affiliation{%
  \institution{Cornell University}
  \city{Ithaca}
  \state{NY}
  \country{USA}
}

\author{Julian Bellavita}
\orcid{0000-0003-1375-5720}
\email{jbellavita@cs.cornell.edu}
\affiliation{%
  \institution{Cornell University}
  \city{Ithaca}
  \state{NY}
  \country{USA}
}

\author{Giulia Guidi}
\orcid{0000-0001-8925-3239}
\email{gguidi@cs.cornell.edu}
\affiliation{%
  \institution{Cornell University}
  \city{Ithaca}
  \state{NY}
  \country{USA}
}

\renewcommand{\shortauthors}{Thomas McFarland, Julian Bellavita, and Giulia Guidi}

\input{sections/abstract}




\maketitle

\input{sections/intro.tex}

\input{sections/background}

\input{sections/relatedwork}

\input{sections/methodology}

\input{sections/resdisc}

\begin{acks}
This research used resources of the National Energy Research
Scientific Computing Center, a DOE Office of Science User Facility
supported by the Office of Science of the U.S. Department of Energy
under Contract No. DE-AC02-05CH11231 using NERSC award
ASCR-ERCAP0030076.
\end{acks}

\bibliographystyle{ACM-Reference-Format}
\bibliography{ref}


\end{document}

%% file: sections/abstract.tex
\begin{abstract}

Sparse General Matrix Multiply (SpGEMM) is key for various High-Performance Computing (HPC) applications such as genomics and graph analytics. 
Using the semiring abstraction, many algorithms can be formulated as SpGEMM, allowing redefinition of addition, multiplication, and numeric types. 
Today large input matrices require distributed memory parallelism to avoid disk I/O, and modern HPC machines with GPUs can greatly accelerate linear algebra computation.

In this paper, we implement a GPU-based distributed-memory SpGEMM routine on top of the CombBLAS library. 
Our implementation achieves a speedup of over $2\times$ compared to the CPU-only CombBLAS implementation and up to $3\times$ compared to PETSc for large input matrices.

Furthermore, we note that communication between processes can be optimized by either direct host-to-host or device-to-device communication, depending on the message size. 
To exploit this, we introduce a hybrid communication scheme that dynamically switches data paths depending on the message size, thus improving runtimes in communication-bound scenarios.

\end{abstract}

%% file: sections/intro.tex
\section{Introduction}
Given matrix multiplication underlies a significant part of HPC applications, considerable efforts have been made in the last three decades to optimize dense matrix multiplication on sequential and parallel machines~\cite{van1997summa, goto2008anatomy, irony2004communication}.
However, many scientific and engineering disciplines rely on Sparse General Matrix Multiply (SpGEMM), where two sparse input matrices are multiplied to produce a sparse output matrix.
For example, each graph can be represented as a sparse adjacency matrix, where the entries denote the edges between the nodes~\cite{10.1145/567112.567114}. 
Problems ranging from quantum chemistry to genomics to graph algorithms can be formulated as sparse graphs and thus as adjacency matrices~\cite{6507483, guidi2021parallel,gleinig2022optimizing} and processed with SpGEMM.

The expressive capabilities of SpGEMM are limited by the fixed, precise nature of matrix multiplication (the multiplication, the addition and the type of the entries or nonzeros).
The semiring abstraction enables user-defined redefinition of entry type, multiplication and addition operator~\cite{glazek2002guide}.
Using the semiring abstraction, numerous graph operations such as breadth first search or $k$-edge shortest path can be reduced to a sparse matrix multiplication over a semiring~\cite{10.1145/3446216}.

Graph sizes have increased rapidly in recent years, for example in genomics with new sequencing technologies~\cite{10.1016/j.gpb.2021.08.001}.
The adjacency matrices of these graphs can exceed single-node memory capacities, so distributed memory parallelism is required to avoid I/O performance degradation.
On the one hand, CombBLAS provides a set of linear algebra primitives with semiring support, parallelized over distributed memory~\cite{bulucc2011combinatorial}.
However, CombBLAS does not currently provide GPU parallelism in its routines, with the exception of one application implemented with the library, HipMCL~\cite{selvitopi2020optimizing}, which does not require semiring support.
On the other hand, GALATIC is a library that supports SpGEMM on custom semirings on a single GPU~\cite{Lett_2021}.

In this paper, we show how distributed SpGEMM can be implemented over semirings with GPUs by integrating the GALATIC matrix multiplication library into CombBLAS by replacing the local multiplication kernel with GALATIC.
Our work also describes our empirical discovery of the performance discrepancy between different communication approaches.
Communication between GPUs can be done either directly between GPUs or by copying to the CPU and communicating between CPUs.
Unexpectedly, we found that the faster method depends on the size of the packet sent.
Thus, we use this discovery to design and implement an heuristic communication switching method and show that it can lead to a noticeable speedup.
To date, most libraries that implement SpGEMM via custom semirings are either distributed but do not have GPU support, or they have GPU support but are not distributed.
Our work demonstrates a successful effort to develop a library that provides both. 
Our software is publicly available here: \url{https://github.com/tf-mac/CombBLAS}.
The main contributions of our work are:

\begin{itemize}
 \item Design, implementation, and evaluation of our extension to CombBLAS with the ability to perform large-scale distributed SpGEMM on GPU using semiring abstraction;
 \item Demonstrate the discovery of inequality between two communication approaches in an MPI implementation;
 \item Present an implementation of communication switching based on this discovery and demonstrate the potential acceleration by offloading GPU communication to the CPU.
\end{itemize}

%% file: sections/background.tex
\section{Background}\label{sec:Background}
This section describes distributed Sparse Generalized Matrix Multiplication (SpGEMM) using semiring abstraction, the CombBLAS and GALATIC libraries, and the different MPI (Message Passing Interface) communication methodologies.

\subsection{Distributed SpGEMM}

SpGEMM is a widely-used kernel for multiplying sparse matrices, often found in graph applications where the adjacency matrices consist mainly of zeros.
Dense matrix multiplication is inefficient for sparse matrices, so more memory-efficient, specialized storage formats such as CSR, CSC or DCSC are used for sparse matrices~\cite{dcsr}.

In graph applications, sparse matrices tend to grow larger than the memory capacity of a single node.
The memory requirement is $\mathcal{O}(n^2)$ with respect to the vertices.
Given this order of magnitude, even high-end computing nodes reach their memory capacity and distributed-memory parallelism is required.
In distributed SpGEMM algorithms, multiple processes handle parts of the multiplication and communicate the results together.
In general, local computation within a node is parallelized via OpenMP~\cite{openmp} and communication between nodes is handled via MPI~\cite{walker1996mpi}.

On distributed systems, we execute SpGEMM via the Sparse Scalable Universal Matrix Multiplication Algorithm (Sparse SUMMA)~\cite{SparseSumma, van1997summa}.
This algorithm divides the processes into a 2D process grid.
The CombBLAS implementation of Sparse SUMMA requires a square number of processes, but the algorithm itself does not.
The multiplication can be regarded as a dot product of subranges of the input matrices \A and \B, corresponding to a subrange of the output matrix \C.
The multiplication is performed by multiplying each process simultaneously over its row and its column, each process being the broadcaster and the receiver of a sub-range of the matrix.
The broadcast replicates the relevant local sub-matrix over the entire row or column of the process grid.

\begin{figure}
    \centering
    \includegraphics[width=1.0\linewidth]{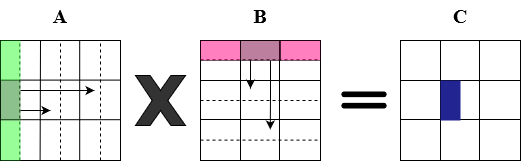}
    \caption{A diagram showing the 2.5D multiplication. Each process column in the \A matrix is split in half, as is each process row in \B. The arrows indicate communication across a processor row or column.}
    \label{fig:2.5D-Comms}
    \vspace{-1em}
\end{figure}

CombBLAS uses a 2.5D algorithm to perform the multiplication.
The $.5$ indicates that each process divides its matrix, resulting in two rounds of multiplication. The \A matrix is halved column-wise, and the \B matrix is halved row-wise.
The actual multiplication is performed by local routines, which reduces the size of the matrices and allows larger matrices to be stored in less memory.
The results are then merged into a product matrix.

\subsection{Semiring Abstraction}

In a nutshell, the semiring abstraction enables the user to redefine the type of matrix entries, multiplication and addition operator.
By definition, a semiring is an algebraic structure that does not require additive inverse~\cite{goodman1999semiring}.
This definition provides a minimal set of axioms with which matrix multiplication can still be performed, since it requires only commutative addition, the existence of zero and one, and associative operations.
It only requires commutative addition, the existence of zero and one, and associative operations and allows the user to redefine addition and multiplication operations.

The semiring abstraction makes it possible to, for example, formulate graph algorithms as matrix multiplication.
In the shortest path problem, we recast the multiplication as addition and addition as the $\min$.
Considering the adjacency matrix of any graph $W$, $W^k$ is the shortest path between vertices using $k$ vertices, and $W^n$ is the distance matrix~\cite{10.1145/567112.567114}.

\subsection{CombBLAS and GALATIC}


CombBLAS is the primary library used to implement the algorithms in this paper~\cite{bulucc2011combinatorial}.
It provides a high-performance implementation of various sparse linear algebra kernels with support for user-defined semirings.
CombBLAS uses the Compressed Sparse Column (CSC) or Doubly Compressed Sparse Column (DCSC) format to store matrices.
However, CombBLAS does not currently support GPU computation.

In contrast, GALATIC~\cite{Lett_2021} provides local SpGEMM on GPU and support for semiring abstraction but does not implement distributed memory parallelism.
By extending the existing AC SpGEMM library~\cite{acspgemm}, GALATIC achieves a significant speed-up compared to the local SpGEMM algorithms of CombBLAS.
GALATIC uses a four-stage process for its operation.
The process begins by dividing the non-zeros of \A evenly among the blocks of a GPU.
Then, the products are computed, expanded, sorted, and combined on the individual GPU threads.
In the third step, parts of the product matrix are combined.
The rows are then merged across the parts before the matrix is returned.
This division of work makes for an extremely efficient SpGEMM, resulting in an algorithm that is faster than cuSPARSE for highly sparse matrices~\cite{acspgemm}.




\subsection{CUDA-aware MPI}

By default, MPI works on CPU memory and communicates data across processes from host memory to host memory. However, there are also CUDA-aware MPI implementations that are often used in a distributed enviroment.
Using CUDA-aware MPI enables direct communication from one GPU memory to another, avoiding the overhead associated with \textsc{cudaMemcpy} by allowing GPUs direct access to each other's memory.

\subsection{CSR, CSC, and DCSC}

Up to three matrix formats are used in the implementation of our SpGEMM routine.
CSC is a column compression format for sparse matrices, where entries are ordered by their column, with their row stored in a separate array to identify their position, and an array for the value of each non-zero entry.
CSR is similar, but instead of columns, the entries are first organized by rows.
DCSC is a variant of CSC that was developed for highly sparse matrices.
It removes completely empty columns and thus reduces the size of the column array.
However, an additional array is required to keep track of the column IDs, unlike normal CSC where the column ID is simply the position in the array.
CombBLAS uses DCSC and CSC, while GALATIC requires CSR. In order to use GALATIC in CombBLAS, a conversion from CSC (or DCSC) to CSR must therefore be carried out.


%% file: sections/relatedwork.tex
\section{Related Work}\label{sec:Related-Work}

Bulu\c{c} et al.~\cite{bulucc2012parallel} developed and compared several distributed SpGEMM algorithms, including the Sparse SUMMA algorithm used in this work.
Azad et al.~\cite{azad2016exploiting} presented a communication avoiding distributed SpGEMM algorithm that uses a 3D process grid and has a lower communication cost compared to Sparse SUMMA.
Hong et al. \cite{hong2024sparsity} presented a sparsity-aware distributed SpGEMM algorithm that reduces the communication cost by communicating only the non-zeros involved in the computation.
These works do not consider the use of GPUs or explore hybrid communication.

Brock et al.~\cite{brock2024rdma} present a number of RDMA-based algorithms for distributed SpGEMM on GPUs, but do not investigate arbitrary semirings or hybrid communication.
A large body of existing work~\cite{dalton2015optimizing, nagasaka2017high, acspgemm} focuses on optimizing SpGEMM on single GPUs.
Other works~\cite{park2023orchestrating, liu2015framework} present strategies for using CPUs and GPUs on the same node in lockstep to perform SpGEMM. However, these works do not target distributed or multi-GPU scenarios.

%% file: sections/methodology.tex
\section{Proposed Implementation}

In this section, we describe our approach to extending the existing CombBLAS by adding the ability to perform large-scale distributed SpGEMM on GPU using semiring abstraction and we show how hybrid communication can be implemented to improve performance.


\subsection{Preparation Phase}

GALATIC and CombBLAS use two different matrix storage format.
GALATIC primarily uses the CSR format, while CombBLAS uses either the CSC or DCSC format, whereby the matrix in the former is organized in a row major format, while in the latter two it is in column major.
A transition to and from each format is required.

For general matrices, the equality $AB = (B^T A^T)^T$ is useful. 
By reinterpreting the column pointer array of CSC as the row pointer of CSR and swapping the row index array, we obtain the transpose of the original matrix in CSR format. 
The same applies to the reverse process.
Our algorithm constructs the final output matrix by tupling, where transposing is equivalent to swapping two entries in each tuple. This method efficiently converts between CSR and CSC formats without recomputing the arrays.

The inequality is only valid for semirings with commutative multiplication. Therefore, we limit ourselves to commutative semirings in this paper and leave the generalization to future work.
The restriction can be circumvented for non-commutative semirings: If you swap the elements of the multiplication (i.e., $a + b \to b + a$), the transposition trick can lead to correct results.

GALATIC requires matrices in GPU memory for local multiplication.
Since CombBLAS stores matrices in CPU memory, we first copy them to GPU memory, where the transpose trick is performed.
DCSC matrices must be converted to CSC so that the transpose can be performed trivially.
For this purpose, the array is rebuilt with the column pointers without skipping additional columns.
To achieve this goal, we ``decompress'' the column pointer array by iterating through the already compressed column pointer array and adding empty segments to the decompressed column pointer array where there were empty columns in the original array.
The pseudocode representing the three steps for preparing the matrices for SpGEMM can be seen in Algorithm \ref{alg:prep}.


\begin{algorithm}[t!]
\caption{Matrix Preparation for SpGEMM}\label{alg:prep}
\begin{algorithmic}[1]
\STATE \textbf{Input:} Matrix \A
\STATE \textbf{Output:} Prepared matrix in GPU memory
\IF{\A is in DCSC format}
    \FOR{Column : compressedColumns}
    \WHILE{isEmpty(Column)}
        \STATE addColumn(col\_ids)
    \ENDWHILE
    \ENDFOR
\ENDIF
\STATE \textbf{Transpose \A}
\STATE \quad Row Ids $\leftrightarrow$ Column Ids
\STATE \textbf{Allocate GPU memory for \A}
\STATE \quad  $\texttt{cudaMalloc}$(\A.row\_pointers[])
\STATE \quad $\texttt{cudaMalloc}$(\A.col\_ids[])
\STATE \quad $\texttt{cudaMalloc}$(\A.data[])
\STATE \textbf{Copy \A from CPU to GPU memory}
\STATE \quad $\texttt{cudaMemcpy}$(\A.row\_pointers[], host$\to$device)
\STATE \quad $\texttt{cudaMemcpy}$(\A.col\_ids[], host$\to$device)
\STATE \quad $\texttt{cudaMemcpy}$(\A.data[], host$\to$device)
\end{algorithmic}
\end{algorithm}

To minimize transposes and copies on the GPU, the preparation is performed at the beginning of the SpGEMM routine.
Only the new local matrix in GPU memory is used, while the original CPU matrix is not reused. Henceforth, we assume that the matrix is located in the GPU memory.

\subsection{Communication}

In this paper, we use the 2.5D communication pattern to limit memory usage, following the CombBLAS communication pattern. Figure~\ref{fig:2.5D-Comms} outlines the 2.5D process.
By keeping the CombBLAS framework, we adapted it for GPU implementation with GPU-aware MPI broadcast.


The local SpGEMM of each sub-matrix is more complicated.
CombBLAS divides the dimensions and the number of nonzeros of each sub-matrix before each local multiplication with each process.
To determine the message sizes, we use the sizes of each sub-matrix that has already been communicated.
If the process is not the root, a new matrix is assigned to receive the message.
 Communication about the process row or column is done via \texttt{MPI\_Bcast}, where the sizes of the sub-matrix are known.

\begin{figure}
    \centering
    \includegraphics[width=.8\linewidth]{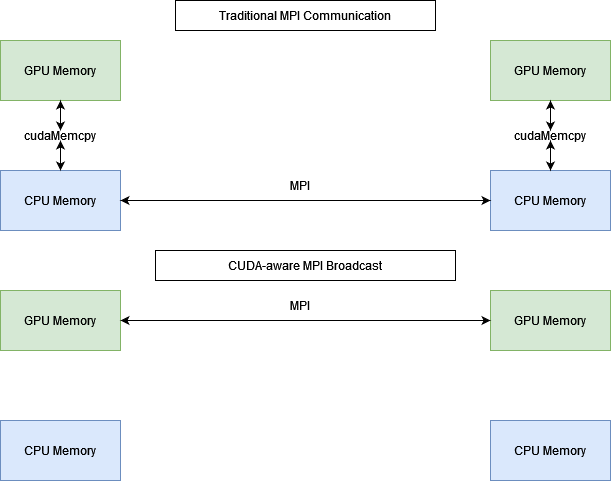}
    \caption{Difference between traditional MPI GPU-GPU communication and CUDA-aware communication.
    }
    \vspace{-1em}
    \label{fig:cuda-aware}
\end{figure}

In our first implementation attempt, inter-process communication involved copying data back to the CPU, broadcasting between host memories, and copying back to GPU memory, which was inefficient.
By using CUDA-aware MPI, this overhead can be avoided, as can be seen in Figure \ref{fig:cuda-aware}. 
CUDA-aware MPI enables direct device-to-device communication without intermediary host copies.
The overhead when copying between CPU and GPU surpasses CPU-CPU communication speed only above a certain, empirically determined message size.
CUDA-aware MPI is therefore only used above this size, which can be configured as an optional parameter.
Below this threshold, traditional CPU-CPU communication is used, as shown in Figure \ref{fig:cuda-aware}.

\subsection{GALATIC SpGEMM}


Once the preparation is complete, the multiplication routine is called.
GALATIC is highly configurable and has parameters for threads, threads per nonzero, and other ratios.
In general, we have found that tuning these parameters does not have a significant impact on performance, but above a certain size and nonzero density, crashes can occur if these parameters are not set.
The \textsc{MaxChunksGeneralizedMerge} setting has been adjusted to handle denser matrices by doubling the allowed matrix density for some shared memory segments. The configuration can be customized to the user's need, e.g., performance or other behavior.
If both input matrices have at least one nonzero, the multiplication \B$^T$\A$^T$ is performed.
The resulting matrix is copied to the CPU, the GPU matrix is freed, and the CPU matrix is returned.



\subsection{Merging Operation}\label{sec:Merging}

Once the local SpGEMMs have been executed, the output matrices are converted to COO format (a list of tuples containing the row, column, and value of each element).
This is done on the CPU.
The transpose is performed by swapping the row and column of each element at the end.
CombBLAS can natively convert this type into a sparse matrix, which is returned at the end of the method.
In future work, we will investigate the implementation of tupling and merging on the GPU.



%% file: sections/resdisc.tex
 \section{Results and Discussion}

The experiments were performed on the GPU nodes of the Perlmutter Supercomputer at NERSC~\cite{nersc} using Cray-MPICH and CUDA Toolkit 12.4.
Each GPU node is equipped with an AMD EPYC 7763 CPU, four NVIDIA A100 GPUs and four Cassini-1 Network Interface Cards (NICs).
The GPUs on the same node are fully connected via 4 NVLink 3.0. The different nodes are connected via a 3-hop Dragonfly topology with an HPE Slingshot 11 interconnect.

The GPUs and CPUs are warmed up with an algorithm run that is not included in the statistics.
I/O is excluded from the specified runtimes.
The values for the first run can be found in Table \ref{tab:first-iter}.
In GALATIC we used the default settings, except for \textsc{MaxChunksGeneralizedMerge}, which was set to 512 instead of 256 to handle larger matrices.
The tests were performed with different matrices from the SuiteSparse Matrix Collection~\cite{davis2019algorithm}, where each operation was a multiplication of a matrix by itself over the semiring of float types.
Relevant statistics can be found in Table \ref{tab:matrix_stats}.

\subsection{Distributed GPU-Enabled SpGEMM}

\input{sections/graphs}
In this section, we evaluate the GPU-enabled SpGEMM in CombBLAS and compare the runtimes with the CPU implementation by measuring the strong scaling performance when increasing the number of processes for different matrix sizes.
To ensure that GPUs were assigned to processes on the same node, subtype communication grids were created for each active node, with processes sharing GPUs.
These grids were for this purpose only and did not affect the overall communication scheme.

For each of the four matrices, we performed three experiments, each of which was repeated four times across four different process grid sizes.
For external comparison, we used the PETSc library\cite{petsc-user-ref}.
PETSc offers 1D SpGEMM with row-wise partitioning and GPU acceleration.
For comparison, we tested each matrix using CombBLAS CPU, CombBLAS GPU (this work), and PETSc GPU and performed $\mathbf{A}^2$ for each matrix.
The tests ran on four nodes, with one GPU per MPI process.
Our GPU routine used only CUDA-aware MPI broadcasts during these tests.
The PETSc results in Figure \ref{fig:long} and the one process experiment in Figure \ref{fig:delan} were excluded due to out of memory issues with large matrices.
Our GPU-enabled CombBLAS version showed a significant speedup, although the magnitude varied depending on the number of processes.
The highest speedup was $2\times$ on the \textsc{Long\_Coup\_dt0} matrix.
Overall, PETSc was faster with smaller matrices, but is slower or unable to run on larger matrices.

In the Figure \ref{fig:atmo} the concept of ``optimal'' grid becomes clear, as we see that the runtimes decrease up to four processes and then increase when the process grid surpasses the optimal size.
The plot shows that the slowdown is due to both the increased communication overhead and the increased pre- and post-processing overhead, but not to the local SpGEMM routine.
This overhead is mainly due to the combination of a 2.5D approach and the CSC format, because to split each matrix in each process, we need to transpose the \B matrix (since it must be split row-wise), which causes a non-trivial overhead now that our SpGEMM routine is optimized.

In addition, we investigated how a different semiring would affect performance.
Figure \ref{fig:semiring} illustrates the \textsc{Long\_dt\_} \textsc{Coup0} matrix under the normal float semiring and the min-plus semiring where the addition is the minimum of the two values, while the multiplication is a normal addition.
The use of simple semirings causes minimal performance losses compared to normal float addition and multiplication.

\subsection{Hybrid Communication}


In our implementation of GPU-GPU communication in CombBLAS, our experiments showed a performance difference between CPU-CPU and GPU-GPU communication.
To evaluate this difference, we compared the latency of \texttt{MPI\_Bcast} with different message sizes using a GPU-GPU communication versus a GPU$\to$CPU copy, a CPU-CPU communication followed by a CPU$\to$GPU copy.
This experiment was performed on 1 node and 4 nodes, as shown in Figure \ref{fig:packet-vs-time}.
This figure shows that, below a certain threshold, broadcast involving direct CPU-CPU communication has lower latency than GPU-GPU communication enabled by CUDA-aware MPI.
This result motivated the introduction of our hybrid communication heuristic based on message size.

\input{sections/comm-graphs}

To determine the appropriate communication heuristic size, we empirically evaluate several threshold sizes.
In these experiments, messages larger than the threshold are sent via GPU-GPU broadcast, while smaller messages are sent via host-memory broadcast.
Our distributed SpGEMM routine was run on the \textsc{rmat} and \textsc{atmosmodd} matrices with different threshold sizes. 
In Figures ~\ref{fig:rmat-comm} and \ref{fig:atmosmodd-comm}, the x-axis is the percentage of broadcasts processed by the CPU compared to the percentage of broadcasts processed by the GPU.
The left y-axis indicates the runtime and the right y-axis the size of the threshold.
Both Figure \ref{fig:rmat-comm} and Figure \ref{fig:atmosmodd-comm} show how the broadcast latency decreases as the threshold increases.
It is worth noting that the runtime and communication time decrease as more of the broadcasts utilize CPU-CPU communication, despite the overhead from the additional memory movement required between CPU and GPU.

In future work, we plan to investigate whether this phenomenon is specific to Perlmutter or whether it also applies to other HPC machines and what the architectural causes of this result might be.
Currently, we hypothesize that this could be due to the overhead of launching the CUDA kernel, as CPU-initiated copy and broadcast operations are almost instantaneous, while GPU-initiated broadcasts may incur additional overhead.
Furthermore, in the future we plan to explore a heterogeneous communication- computation mapping, where both the local SpGEMM and the communication between processes are mapped to either the CPU or the GPU.
This is a challenging optimization problem, as one must take into account that copying to the GPU may be delayed or non-existent in the receiving node.
That is, there may be cases where GPU-GPU communication would be optimal if the receiving node is computing on the GPU, but suboptimal if the receiving node is computing on the CPU.
this is a complex optimization problem and an interesting future research direction.

\section{Conclusions}\label{sec:conc-futur}

SpGEMM is a crucial kernel for solving high-performance computing challenges, and several applications in the sciences can be reduced to sparse matrix multiplication over a user-defined semiring.
The matrices often grow beyond the memory and computational capacities of individual nodes.
Therefore, distributed-memory parallel sparse matrix multiplication is required to solve these challenges efficiently.

In this paper, we presented a scalable GPU-enabled, distributed SpGEMM routine that support an arbitrary semiring.
To achieve this goal, we integrated the GALATIC library into CombBLAS to effectively handle local SpGEMMs on the GPU.
In addition, we modified the communication routines to operate on matrices resident in GPU memory.
In this way, we were able to achieve a 1.5-2.5$\times$ speedup compared to the native CPU version of CombBLAS and up to 3$\times$ speedup compared to the GPU SpGEMM routine in PETSc.


Finally, we showed that inter-process communication can be faster when copied to the CPU and CPU-CPU communication is used instead of direct GPU-GPU communication via CUDA-aware MPI.
Based on this insight, we implemented a hybrid communication scheme based on message sizes.
This resulted in a speedup of the communication runtime between 2$\times$ and 4$\times$.

%% file: sections/graphs.tex
\begin{table}[t]
\centering
\caption{Runtimes of the first iteration (s) of CombBLAS on 4 nodes of Perlmutter varying the number of processes (P).}
\label{tab:first-iter}
\footnotesize
\begin{tabularx}{\columnwidth}{c|>{\centering\arraybackslash}X>{\centering\arraybackslash}X|>{\centering\arraybackslash}X>{\centering\arraybackslash}X|>{\centering\arraybackslash}X>{\centering\arraybackslash}X|>{\centering\arraybackslash}X>{\centering\arraybackslash}X}
\toprule
P & \multicolumn{2}{c}{rmat} & \multicolumn{2}{c}{atmosmodd} & \multicolumn{2}{c}{delaunay} & \multicolumn{2}{c}{Long\_dt\_Coup0} \\
\cmidrule(r){2-3} \cmidrule(r){4-5} \cmidrule(r){6-7} \cmidrule(r){8-9}
 & CPU & GPU & CPU & GPU & CPU & GPU & CPU & GPU \\
\midrule
1 & 4.19 & 2.61 & 3.62 & 3.66 & 36.37 & 23.46 & 61.34 & 37.29 \\
4 & 1.27 & 0.85 & 2.14 & 1.82 & 17.02 & 9.14 & 37.02 & 19.79 \\
9 & 3.20 & 2.79 & 5.42 & 4.77 & 27.71 & 13.91 & 78.92 & 40.67 \\
16 & 3.64 & 3.90 & 7.21 & 6.02 & 41.85 & 18.94 & 103.36 & 43.53 \\
\bottomrule
\end{tabularx}
\end{table}

\begin{table}[t]
    \centering
    \caption{List of matrices used in the evaluation and associated statistics.}
    \begin{tabularx}{\columnwidth}{l>{\centering\arraybackslash}X>{\centering\arraybackslash}X>{\centering\arraybackslash}X}
        \toprule
        Matrix Name      & Rows     & Columns   & Nonzeros   \\ \midrule
        rmat             & \hspace{1em}65536    & \hspace{1em}65536     & \hspace{1em}490228     \\
        atmosmodd       & 1270432  & 1270432   & \hspace{.5em}8814880    \\
        delaunay\_n22   & 4194304  & 4194304   & 25165738   \\
        Long\_dt\_Coup0  & 1470152  & 1470152   & 70219816   \\ \bottomrule
    \end{tabularx}
    \label{tab:matrix_stats}
\end{table}

\begin{figure}
    \centering
    \includegraphics[width=1\linewidth]{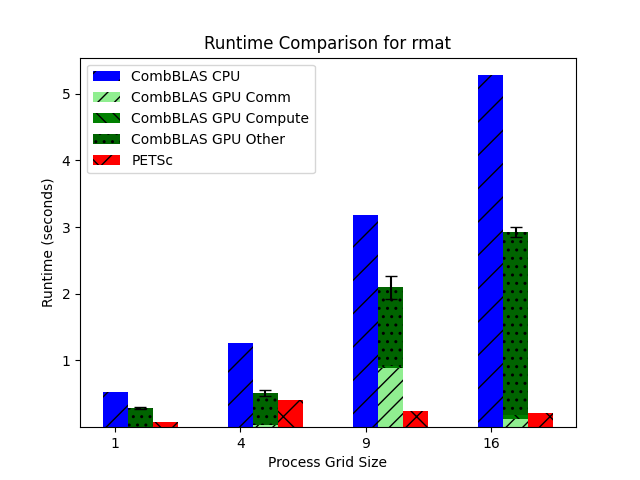}
    \caption{Runtime of the \textsc{rmat} matrix under CombBLAS CPU, GPU, and PETSc.}
    \label{fig:rmat}
\end{figure}

\begin{figure}
    \centering
    \includegraphics[width=1\linewidth]{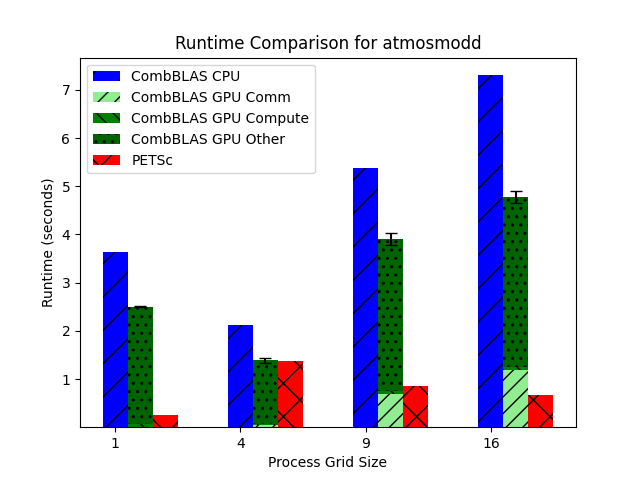}
    \caption{Runtime of the \textsc{atmosmodd} matrix under CombBLAS CPU, GPU, and PETSc.}
    \label{fig:atmo}
\end{figure}

\begin{figure}
    \centering
    \includegraphics[width=1\linewidth]{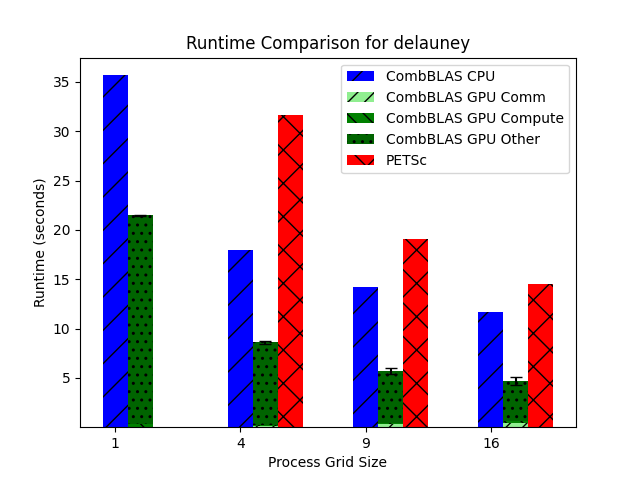}
    \caption{Runtime of the \textsc{delaunay} matrix under CombBLAS CPU, GPU, and PETSc.}
    \label{fig:delan}
\end{figure}

\begin{figure}
    \centering
    \includegraphics[width=1\linewidth]{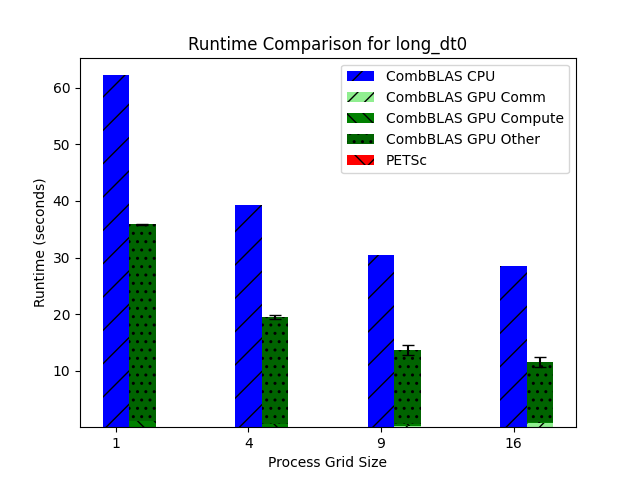}
    \caption{Runtime of the \textsc{LongDt0} matrix under CombBLAS CPU and GPU.}
    \vspace{-1em}
    \label{fig:long}
\end{figure}
\begin{figure}
    \centering
    \includegraphics[width=1\linewidth]{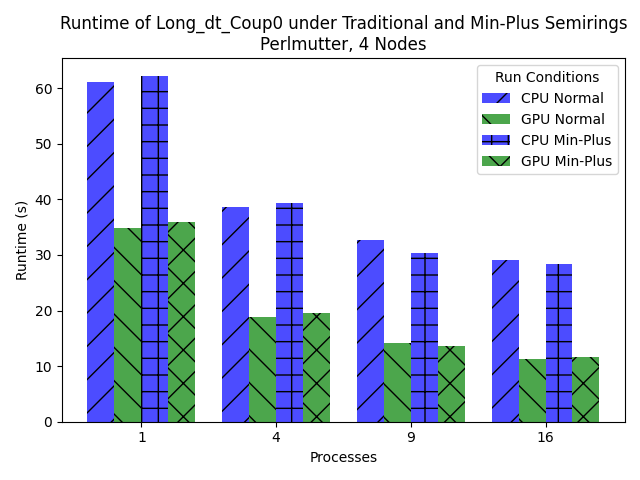}
    \caption{Runtime of the \textsc{LongDt0} matrix under CombBLAS CPU, GPU, and a Min-Select Semiring. 
    }
    \vspace{-1em}
    \label{fig:semiring}
\end{figure}
\begin{figure}
    \centering
    \includegraphics[width=1\linewidth]{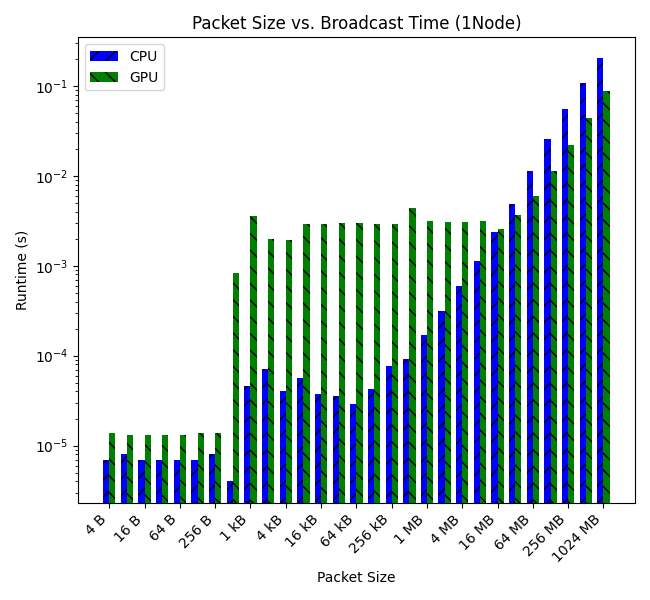}
    \includegraphics[width=1\linewidth]{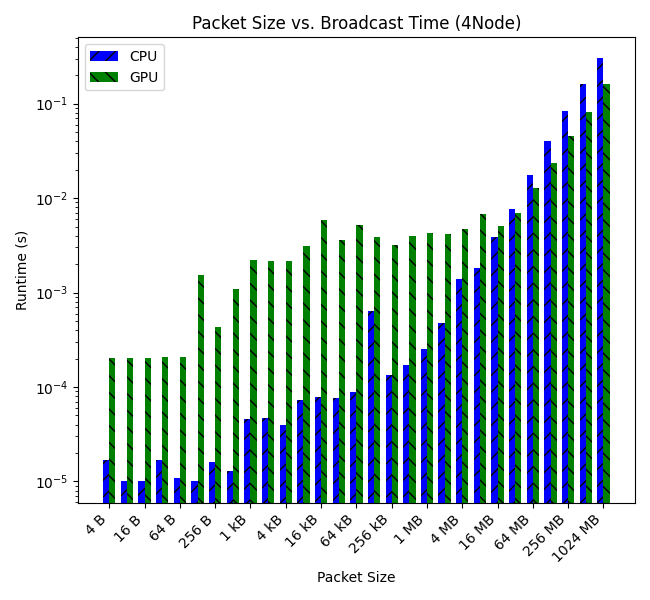}
    \caption{Packet Size vs. Broadcast Time on a logarithmic scale for 1 Node (top) and 4 Nodes (bottom), using the two communication methodologies.}

    \label{fig:packet-vs-time}
\end{figure}

%% file: sections/comm-graphs.tex
\begin{figure}
    \centering
    \includegraphics[width=\columnwidth]{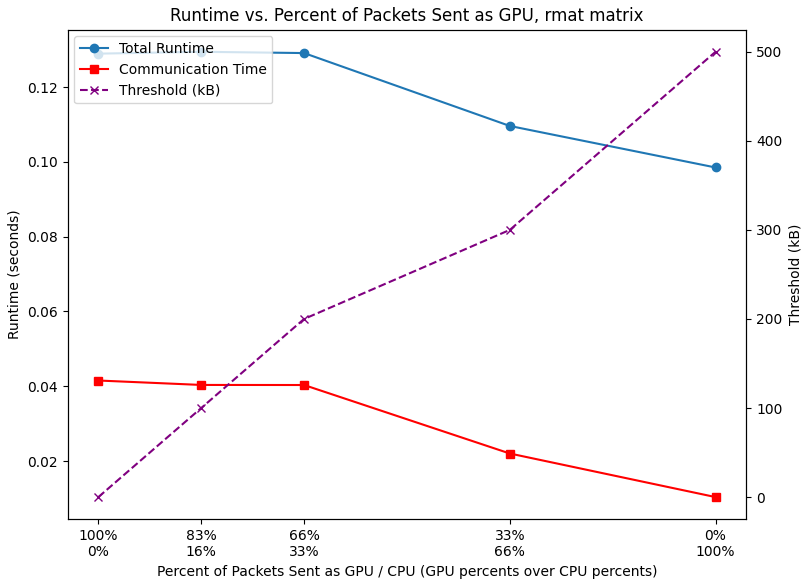}
    \caption{Runtime of the \textsc{rmat} matrix varying the commonication threshold.}
    \label{fig:rmat-comm}
\end{figure}

\begin{figure}
    \centering
    \includegraphics[width=\columnwidth]{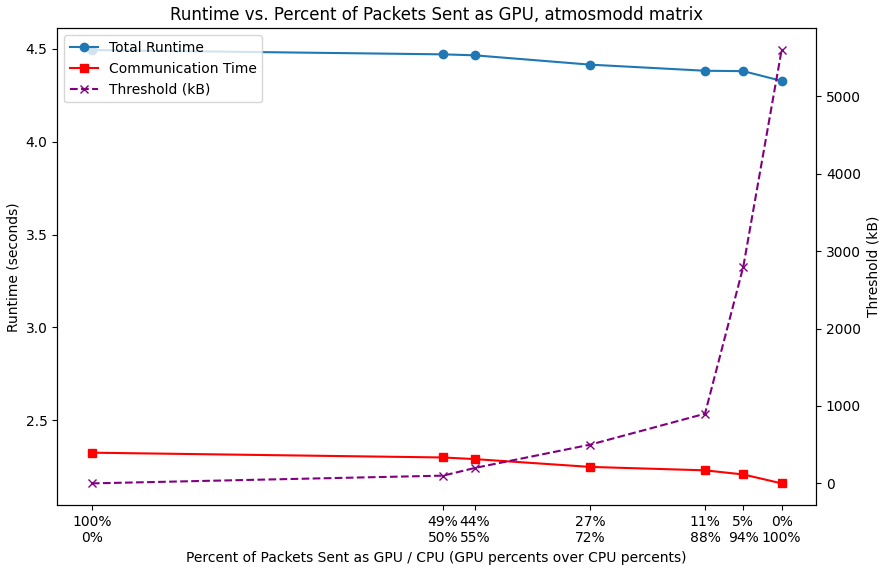}
    \caption{Runtime of the \textsc{atmosmodd} matrix varying the commonication threshold.}
    \label{fig:atmosmodd-comm}
\end{figure}